# Biaxial strain tuning of interlayer excitons in bilayer MoS₂

*Felix Carrascoso[1], Der-Yuh Lin[2], Riccardo Frisenda[1,*], Andres Castellanos-Gomez[1,*]*

[1] *Materials Science Factory, Instituto de Ciencia de Materiales de Madrid (ICMM), Consejo Superior de Investigaciones Científicas (CSIC), Sor Juana Inés de la Cruz 3, 28049 Madrid, Spain.*
[2] *National Changhua University of Education, Bao-Shan Campus, No. 2, Shi-Da Rd, Changhua City 500, Taiwan, R.O.C*

\* *riccardo.frisenda@csic.es , andres.castellanos@csic.es*

## ABSTRACT

We show how the excitonic features of biaxial MoS₂ flakes are very sensitive to biaxial strain. We find a lower bound for the gauge factors of the A exciton and B exciton of (-41 ± 2) meV/% and (-45 ± 2) meV/% respectively, which are larger than those found for single-layer MoS₂. Interestingly, the interlayer exciton feature also shifts upon biaxial strain but with a gauge factor that is systematically larger than that found for the A exciton, (-48 ± 4) meV/%. We attribute this larger gauge factor for the interlayer exciton to the strain tunable van der Waals interaction due to the Poisson effect (the interlayer distance changes upon biaxial strain).

The isolation of atomically thin $MoS_2$ by mechanical exfoliation in 2010 opened the door to study the intriguing optical properties of this 2D semiconductor material. [1–3] In fact, $MoS_2$ and other members of the transition metal dichalcogenide family show a rich plethora of exitonic physical phenomena, present even at room temperature. Mak *et al.* and Splendiani *et al.* observed a strong thickness dependent photoluminescence emission and a direct-to-indirect band gap transition.[1,2] $MoS_2$ also has tightly bound negative trions, an exciton quasiparticle composed of two electrons and a hole,[4,5] and several groups in parallel reported the valley polarization, the selective population of one valley, by pumping with circularly polarized light.[6–8] Moreover, heterostructures built with these 2D systems have also attracted the interest of the scientific community because of the presence of interlayer excitons: excitons formed by electrons and holes that live in different layers.[9–13] Very recently, Gerber *et al.* and Slobodeniuk *et al.* demonstrated that naturally stacked bilayer $MoS_2$ (2H- polytype) also presents interlayers excitons, with high binding energy, that can be observed at room temperature [14,15] and Niehues *et al.* demonstrated that uniaxial strain could be used to tune the energy of the interlayer exciton.[16]

In this work we employ biaxial strain to modify the band structure, and thus the excitonic resonances, in bilayer $MoS_2$ flakes. We observe that both the A and B excitons, as well as the interlayer exciton, substantially redshift upon biaxial tension. Interestingly, unlike to what has been reported for uniaxial strain, we found that the interlayer exciton is more effectively tuned upon straining than the A and B excitons. We attribute this effect to a modification of the interlayer interaction as an in-plane biaxial expansion of the bilayer $MoS_2$ is expected to come hand-by-hand of an out-of-plane compression due to the $MoS_2$ Poisson's ratio.

MoS₂ flakes were prepared by mechanical exfoliation of bulk natural molybdenite (Moly Hill mine, QC, Canada) with Nitto tape (SPV 224). The cleaved MoS₂ flakes are then transferred to a Gel-Film (WF 4x 6.0 mil Gel-Film from Gel-Pak®, Hayward, CA, USA). The flakes are optically identified, and their number of layers are determined from quantitative analysis of transmission mode optical microscopy images and micro-transmittance/reflectance spectroscopy.[17–19] Once a suitable bilayer MoS₂ flake is located it is transferred onto a polypropylene (PP) substrate by an all-dry deterministic transfer method.[20,21]

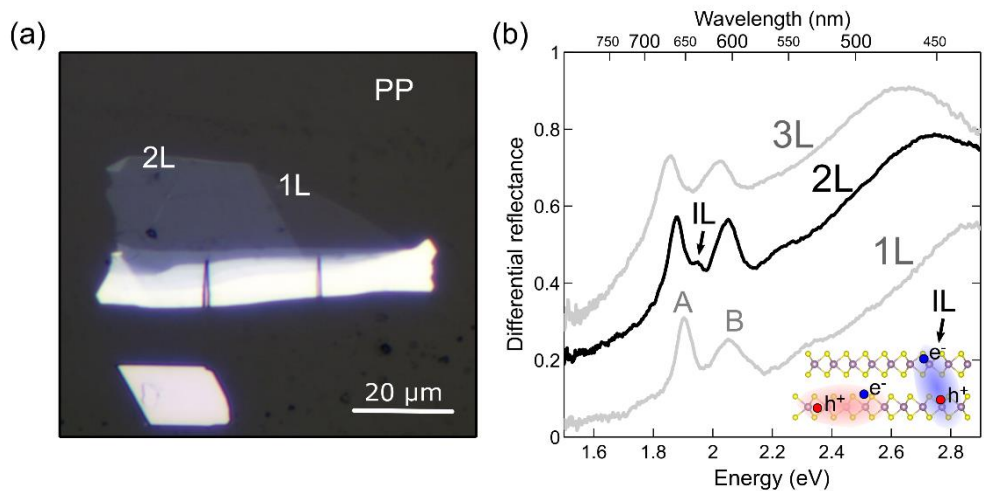

**Figure 1.** (a) Reflection mode optical microscopy image of a MoS₂ flake with single- and bi-layer regions (highlighted with 1L and 2L respectively) transferred onto the surface of a polypropylene (PP) substrate. (b) Differential reflectance spectra acquired on single-, bi- and tri-layer MoS₂ flakes on PP. The A and B excitons are clearly visible for all these thicknesses while the interlayer (IL) exciton, present in multilayer MoS₂, is more clearly visible for the bilayer flake.

Figure 1(a) shows a reflection mode optical microscopy image of a MoS₂ flake transferred onto PP. Figure 1(b) shows differential reflectance spectra acquired on a mono-, bi- and tri-layer MoS₂ flake with a home-built micro-reflectance microscope. We address the reader to Ref. [22] for technical details about the experimental setup. To obtain the differential reflectance spectra we first collect the light reflected from the substrate ($R_s$) by means of a fiber-coupled compact CCD spectrometer (see Materials and Methods). Then we collect the light reflected by the desired MoS$_2$ flake ($R_f$) and we calculate the differential reflectance as: $\Delta R/R = 1 - R_f/R_s$.[19,23] All the spectra displayed in Figure 1(b) show two strong transitions in all of them assigned to the A and B excitons (~1.9 eV and ~2.05 eV respectively) originated from direct band gap transitions at the K point of the Brillouin zone.[1,2] Interestingly, in bilayer MoS$_2$ one can see another prominent peak between the A and B excitons. That peak can be also observed in trilayer and even multilayer MoS$_2$ but it cannot be as easily resolved as in the case of bilayer MoS$_2$. This feature in the reflectance spectra have been recently demonstrated (through temperature dependent optical spectroscopy studies, magneto-optical measurements and density functional theory calculations) to be originated by the generation of interlayer (IL) excitons.[14,15,24] These excitons are, similarly to the A and B excitons, due to direct transitions at the K point but unlike them the electron and hole are spatially separated in the different MoS$_2$ layers (see the cartoon in Figure 1(b)).

Uniaxial and biaxial strain have been proven to be effective methods to modify the optical properties of 2D semiconductors.[25–28] Here, in order to biaxially strain the MoS$_2$ bilayers we exploit the large thermal expansion mismatch between the PP substrate (~130×10$^{-6}$ K$^{-1}$) and MoS$_2$ (1.9×10$^{-6}$ K$^{-1}$)[29]. PP has also a relatively high Young's modulus (1.5-2 GPa) for a polymer, which is essential to guarantee an optimal strain transfer from substrate to flake. One can then biaxially stretch (or compress) the flakes by warming up (or cooling down) the substrate.[30–32] We used a Peltier element to control the temperature of the substrate around room temperature (27-28ºC) that allows us to cool down to 17ºC (-0.13%) and to warm up to 95ºC (+0.87%). The substrate temperature can be translated to biaxial expansion/compression through the thermal expansion coefficient of PP (see the Supporting Information).

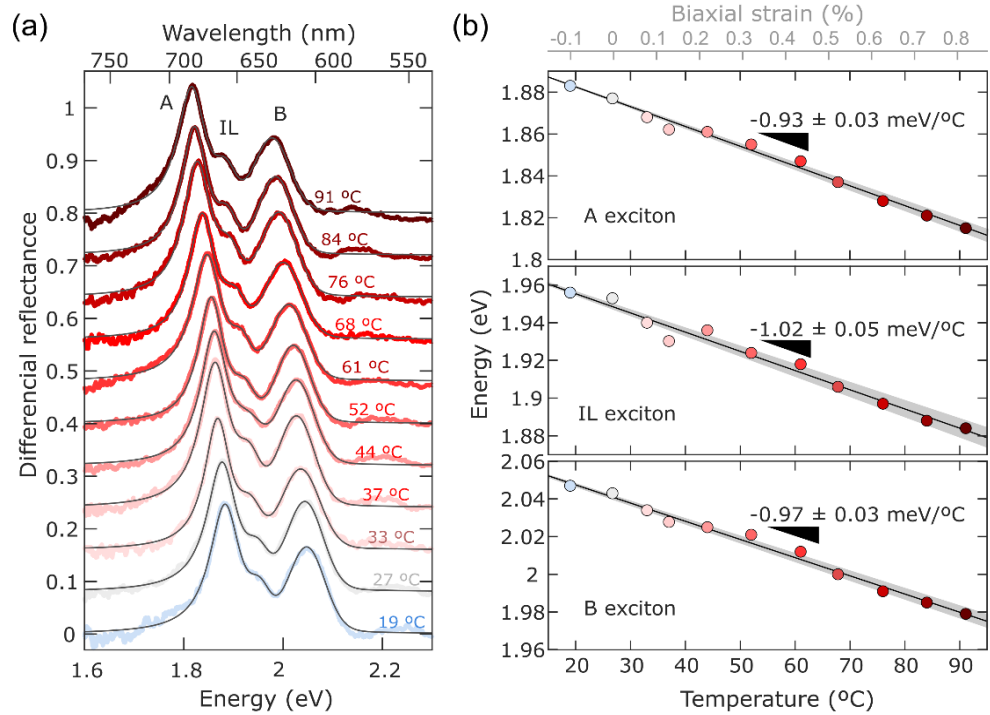

**Figure 2.** (a) Differential reflectance spectra of a MoS$_2$ bilayer deposited on PP recorded at different temperatures (quadratic polynomial background removed). The black solid lines represent the total fit to the data (composed of three Gaussian peaks). (b) energy of the A, B and IL excitonic peaks extracted from the fit and plotted as a function of the substrate temperature (bottom axis) and of the substrate biaxial strain (top axis). Note that the uncertainty of the exciton energies determined through the fits is below 0.1% of their value.

Figure 2(a) shows the differential reflectance spectra acquired at different substrate temperatures from 19°C to 91°C. The spectra redshift upon temperature increase above room temperature and blueshift when the substrate is cooled down below room temperature. The spectra can be fitted to a sum of three Gaussian peaks in order to extract the energy position of the A, B and IL excitons. The summary of the exciton energy positions is shown in Figure 2(b). From this figure one can extract the spectral shift per °C for the different excitons: (-0.93 ± 0.03) meV/°C for the A exciton and (-0.97 ± 0.03) meV/°C for the B exciton. In order to disentangle the intrinsic spectral shift expected for $MoS_2$ upon temperature change from that originated from the biaxial strain we fabricated a $MoS_2$ sample on a Si substrate with 50 nm of $SiO_2$, which is expected to have a negligible thermal expansion, and we probe the exciton position as a function of temperature. We found that for single-, bi- and tri-layer $MoS_2$ all the excitons shift by – 0.4 meV/°C. By subtracting this intrinsic thermal shift value to the values measured in samples fabricated on PP we can determine the spectral shift induced by biaxial strain. And we can determine the gauge factor, the spectral shift per % of biaxial strain, by calculating the substrate biaxial expansion (or compression) upon temperature change (see the Supporting Information for details about the thermal expansion calibration of the PP substrates). The resulting gauge factors for the A and B excitons are (-41 ± 2) meV/% and (-45 ± 2) meV/%. Note that these gauge factor values should be considered as a lower bound as we are assuming that all the biaxial expansion of the substrate can be effectively translated to biaxial strain to the $MoS_2$ flake. Due to the Young's modulus mismatch between the PP substrate and the $MoS_2$ the strain transfer efficiency could be lower (and thus we would be underestimating the gauge factor values).[30,33] The fact that all the spectra shows a clear IL peak indicates that both $MoS_2$ layers are equally strained (maintaining the 2H- stacking during the whole straining cycle) as the presence of IL exciton peaks is extremely sensitive to the relative atomic arrangement between the layers.[14] We also address the reader to the Supporting Information section S6 for a finite element simulation used to estimate the strain transfer along the thickness of thick multilayered $MoS_2$ flakes.

It is interesting to note that all the bilayer $MoS_2$ flakes studied here have A and B exciton gauge factors that are substantially larger than those found for single-layer flakes which are in the –(10-25) meV/% range [30], in agreement with density functional theory calculations [34]. We point the reader to the Supporting Information for a summary of the measured datasets in 2 single-layer flakes, other 5 bilayers and one trilayer flake.

For the interlayer exciton we find a gauge factor of (-48 ± 4) meV/% which is substantially larger than that found for the A exciton. We address the reader to the Supporting Information for datasets acquired on other five bilayer $MoS_2$ flakes (with gauge factor up to -55 meV/%) and one trilayer flake that also have a substantially larger gauge factor for the interlayer exciton (-23 meV/%) than for the A exciton (-11 meV/%) similarly to the bilayer case. This contrasts with what has been recently reported for uniaxially strained bilayer $MoS_2$ flakes by Niehues *et al.* where the gauge factor of the interlayer exciton was slightly lower than that of the A exciton. We attribute the larger gauge factor observed in our experiment to a reduction (or increase) of the bilayer interlayer spacing upon biaxial tension (or compression) as expected from the Poisson effect: as the out-of-plane Poisson's ratio of $MoS_2$ is $v_o \sim 0.2$ a biaxial tension of 1% would yield a reduction of 0.2% in the interlayer distance.[35] A similar tunability of the interlayer van der Waals interaction upon biaxial strain has been recently reported in black phosphorus by Huang and co-workers.[36] The strain tunable interlayer distance could explain the large gauge factor observed for bilayer $MoS_2$ upon biaxial strain as Deilmann and Thygesen demonstrated through density functional theory calculations that the interlayer exciton position strongly depends on the interlayer distance.[24] In previous uniaxial strain works, on the other hand, because of the Poisson's ratio of the polycarbonate substrate ($v = 0.37$) when the flake is uniaxially stretched in one direction it is compressed in the perpendicular direction (within the basal plane)[37] counteracting most of the Poisson's effect induced upon uniaxial tension.[16]

In Figure 3 we test the reproducibility of the biaxial strain tuning exploiting the thermal expansion of the substrate. We modulated the temperature of the substrate between ~30ºC and ~40ºC (see the registered temperature *vs.* time in the top panel of Figure 3). The color map in the bottom panel shows the time evolution of the differential reflectance spectra and the extracted position of the A, IL and B excitons, extracted from fits similarly to Figure 2(a), are displayed with the black lines. This illustrates the power of this method to tune the van der Waals interlayer interaction and thus the interlayer excitons in biaxial MoS$_2$.

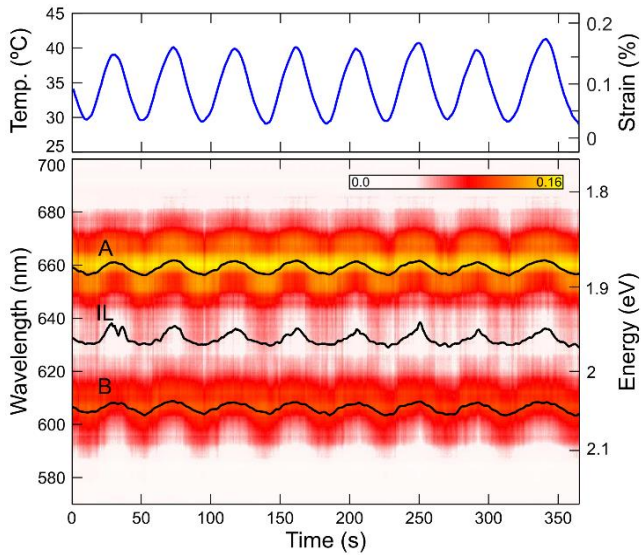

**Figure 3.** Time evolution of the differential reflectance spectra of bilayer MoS$_2$ (bottom axis) while the temperature of the substrate is cycled between 30ºC and 40ºC (top axis). The intensity of the differential reflectance spectra is displayed in the color axis of the colormap. The A, B and IL exciton position is also displayed through the black solid lines.

## CONCLUSIONS

In summary, we have exploited the large thermal expansion of polypropylene substrates to subject biaxial MoS$_2$ flakes to biaxial strain. We find that the excitons redshift upon biaxial tension with gauge factors that are larger than those reported for monolayer MoS$_2$. Interestingly, the interlayer exciton gauge factor is systematically larger than that of the A and B excitons (contrasting the results reported for uniaxially strained bilayer MoS$_2$).

We attribute this larger gauge factor of the interlayer exciton to the strain tuning of the van der Waals interaction upon biaxial in-plane straining due to the Poisson effect.

## ACKNOWLEDGEMENTS

This project has received funding from the European Research Council (ERC) under the European Union's Horizon 2020 research and innovation programme (grant agreement n° 755655, ERC-StG 2017 project 2D-TOPSENSE). ACG acknowledge funding from the EU Graphene Flagship funding (Grant Graphene Core 2, 785219). RF acknowledges support from the Spanish Ministry of Economy, Industry and Competitiveness through a Juan de la Cierva-formación fellowship 2017 FJCI-2017-32919. D.-Y.L. acknowledges the financial support from the Ministry of Science and Technology of Taiwan, Republic of China under contract No. MOST 108-2221-E-018-010.

## MATERIALS AND METHODS

Optical microscopy images have been acquired with an AM Scope BA MET310-T upright metallurgical microscope equipped with an AM Scope MU1803 camera with 18 megapixels. The trinocular of the microscope has been modified to connect it to a fiber-coupled Thorlabs spectrometer (part number: CCS200/M) to perform the differential reflection spectroscopy measurements.[19]

# Supporting Information: Biaxial strain tuning of interlayer excitons in bilayer MoS2

*Felix Carrascoso[1], Der-Yuh Lin[2], Riccardo Frisenda[1,*], Andres Castellanos-Gomez[1,*]*

[1] *Materials Science Factory, Instituto de Ciencia de Materiales de Madrid (ICMM), Consejo Superior de Investigaciones Científicas (CSIC), Sor Juana Inés de la Cruz 3, 28049 Madrid, Spain.*
[2] *National Changhua University of Education, Bao-Shan Campus, No. 2, Shi-Da Rd, Changhua City 500, Taiwan, R.O.C*

* *riccardo.frisenda@csic.es* , *andres.castellanos@csic.es*

**Section S1 – Polypropylene substrate thermal expansion calibration**

**Section S2 – Fitting of the differential reflectance spectra**

**Section S3 – Additional samples**

      Single-layer MoS2 samples

      Bilayer MoS2 samples

      Trilayer MoS2

      Exciton gauge factor statistics for bilayer MoS2

**Section S4 – Disentangling temperature and strain effects**

**Section S5 – Ruling out slippage effects**

**Section S6 – Estimating the strain transfer in multilayered flakes**

## Section S1 – Polypropylene substrate thermal expansion calibration

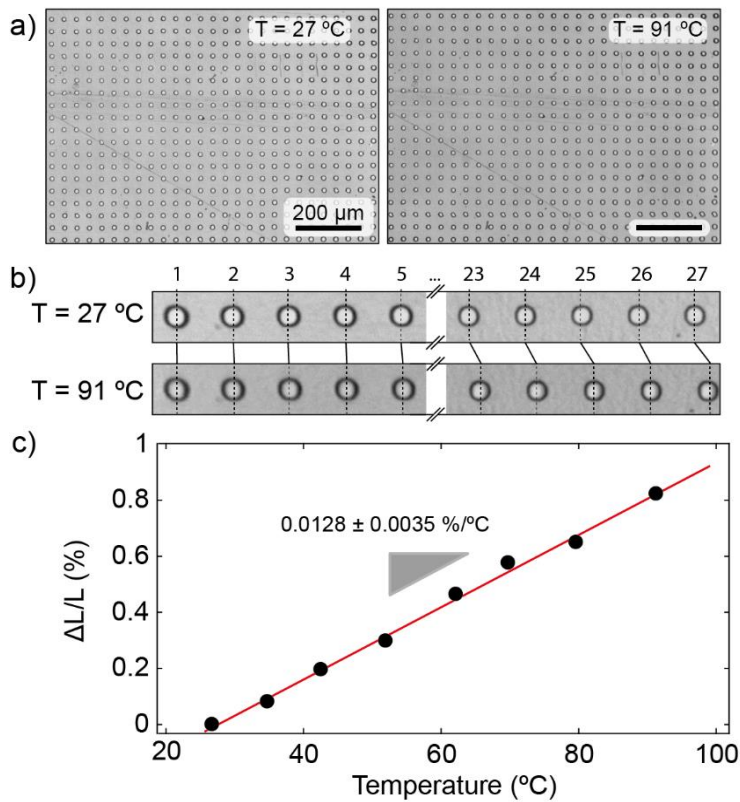

**Figure S1.** (a) Microscope pictures of a PP substrate with a periodic array of pillars defined by optical lithography at two different temperatures. (b) Zoom in on a row of pillars to show the thermal expansion experienced at high temperature. (c) Relative change (ΔL/L) of the spacing between the pillars as a function of temperature. A linear relation can be observed in the experimental temperature range. The thermal expansion coefficient of PP is determined as 128·10⁻⁶ /K from the slope of the linear fitting.

## Section S2 – Fitting of the differential reflectance spectra

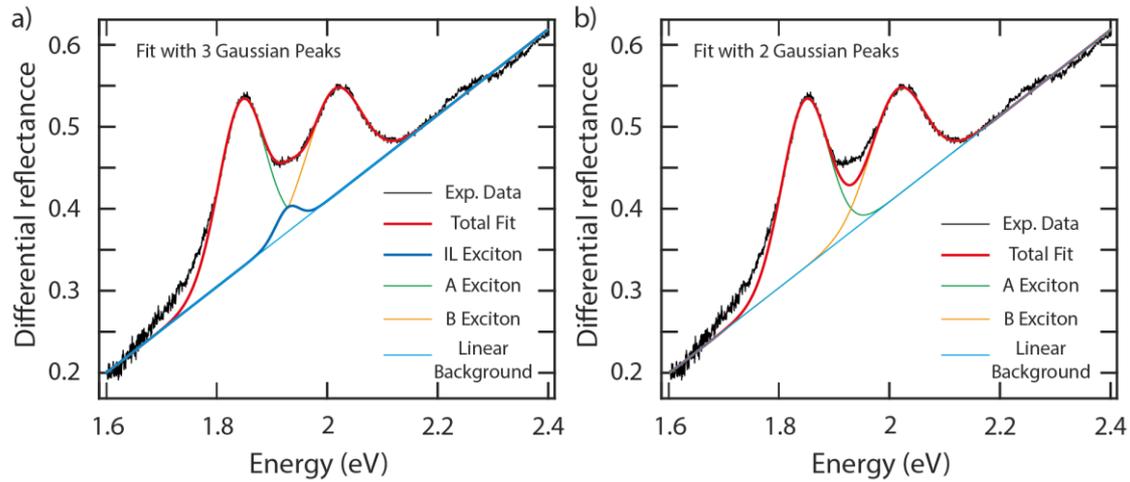

**Figure S2.** (a) Differential reflectance spectrum of a MoS$_2$ trilayer deposited on PP at 27 ºC. The red curve represents the total fit to the data, which is composed by the sum of a linear background and three Gaussian peaks (A, B and IL). (b) Same as (a) with two Gaussian peaks (A and B) instead of three peaks.

## Section S3 – Additional samples

### Single-layer MoS$_2$ samples

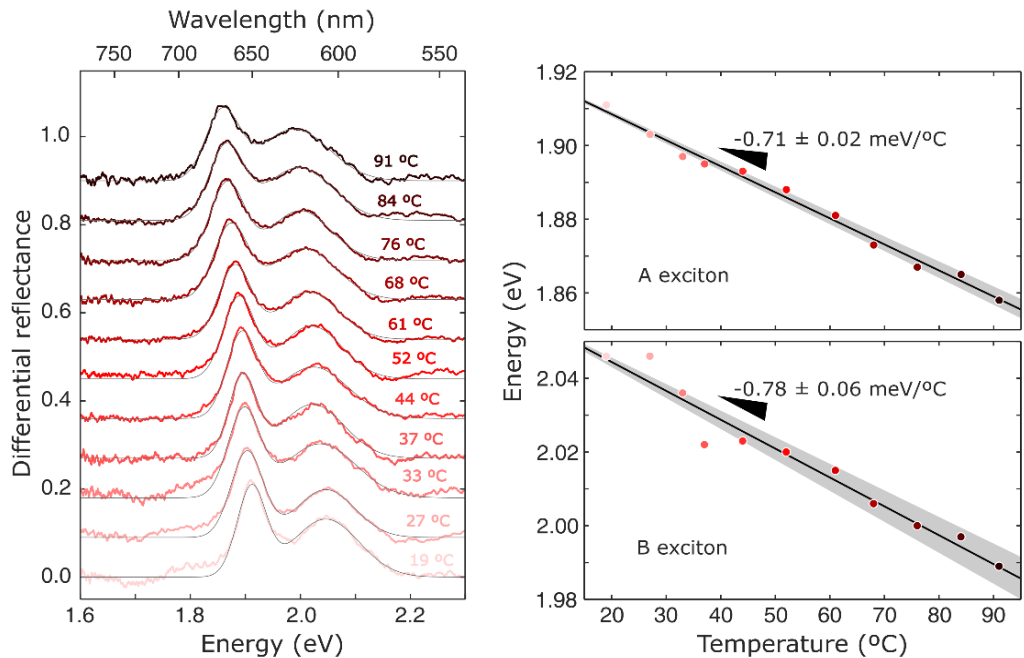

**Figure S3.** Left: differential reflectance spectra of a MoS$_2$ monolayer deposited on PP recorded at different temperatures. The black solid lines represent the total fit to the data (composed of two Gaussian peaks). Right: energy of the A and B excitonic peaks extracted from the fit plotted as a function of the substrate temperature.

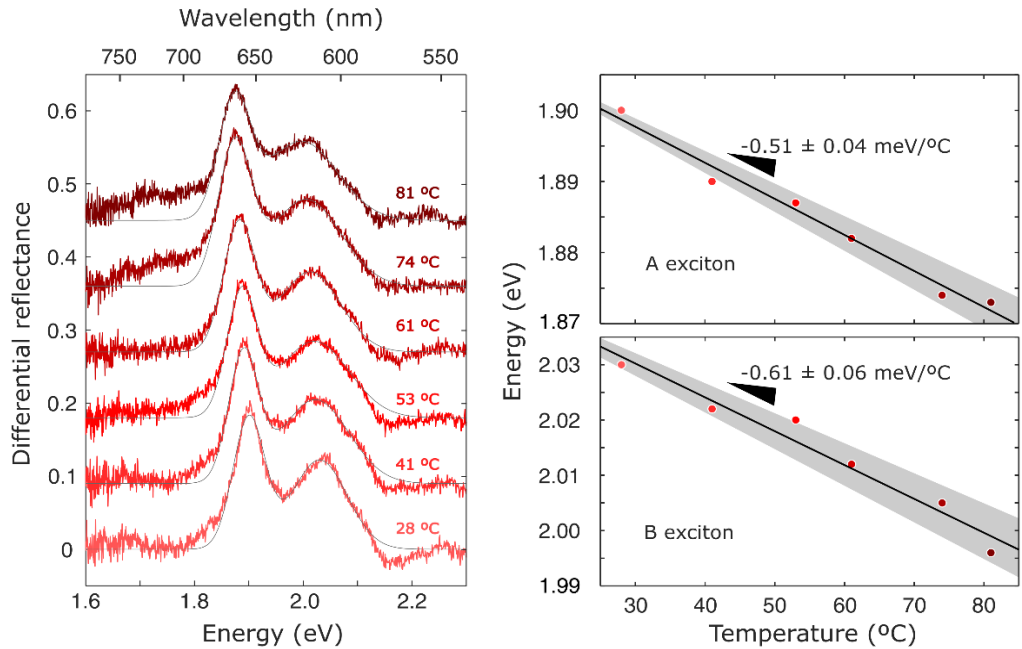

**Figure S4.** Left: differential reflectance spectra of a MoS₂ monolayer deposited on PP recorded at different temperatures. The black solid lines represent the total fit to the data (composed of two Gaussian peaks). Right: energy of the A and B excitonic peaks extracted from the fit plotted as a function of the substrate temperature.

## Bilayer MoS₂ samples

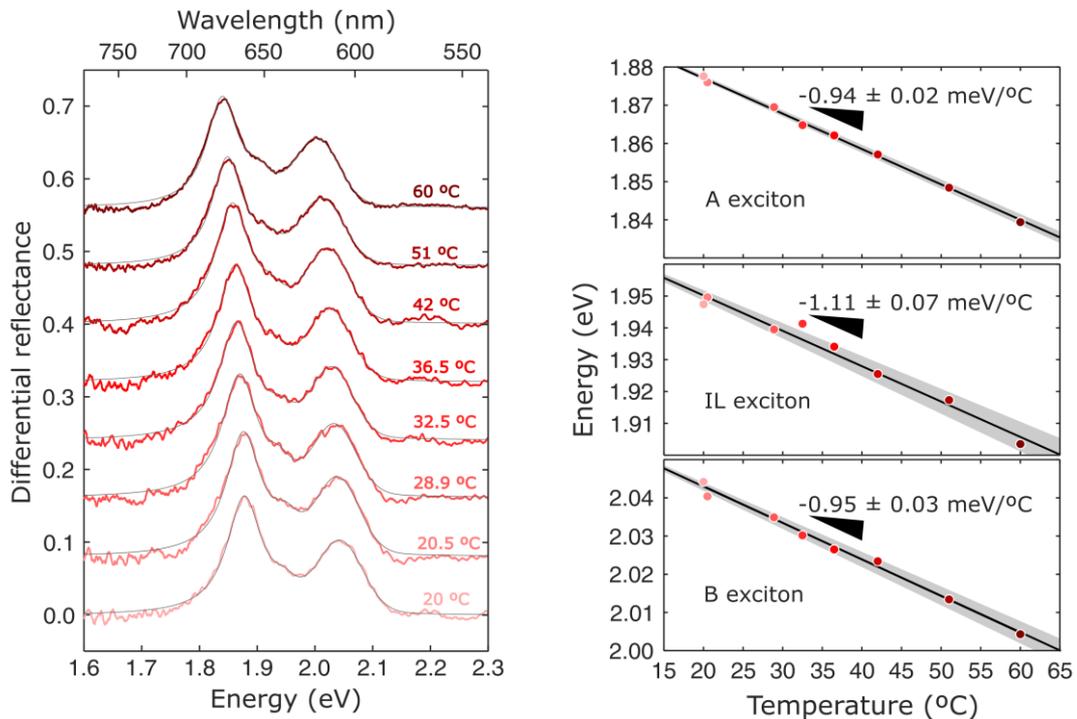

**Figure S5.** Left: differential reflectance spectra of a MoS₂ bilayer deposited on PP recorded at different temperatures. The black solid lines represent the total fit to the data (composed of three Gaussian peaks). Right: energy of the A, B and IL excitonic peaks extracted from the fit plotted as a function of the substrate temperature.

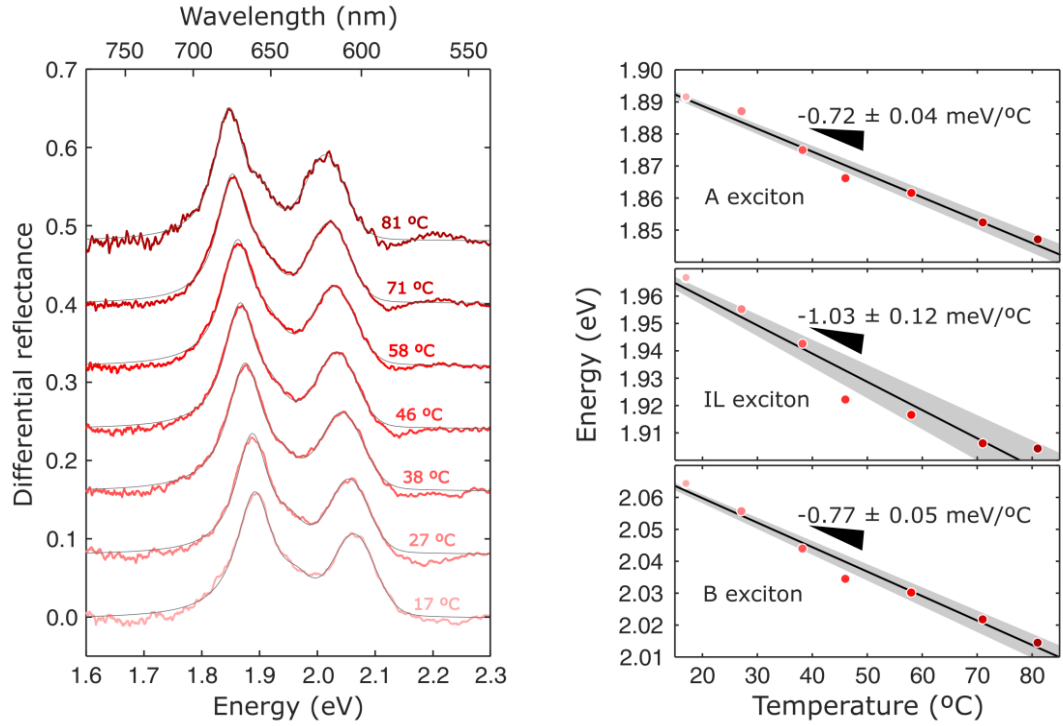

**Figure S6.** Left: differential reflectance spectra of a MoS$_2$ bilayer deposited on PP recorded at different temperatures. The black solid lines represent the total fit to the data (composed of three Gaussian peaks). Right: energy of the A, B and IL excitonic peaks extracted from the fit plotted as a function of the substrate temperature.

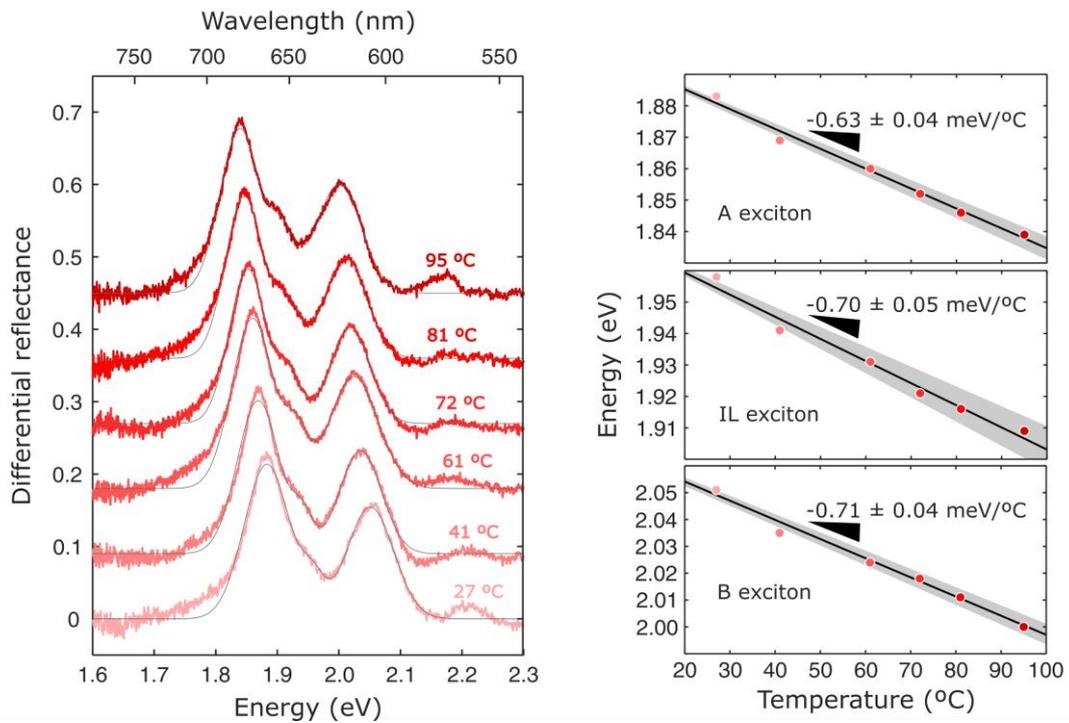

**Figure S7.** Left: differential reflectance spectra of a MoS$_2$ bilayer deposited on PP recorded at different temperatures. The black solid lines represent the total fit to the data (composed of three Gaussian peaks). Right: energy of the A, B and IL excitonic peaks extracted from the fit plotted as a function of the substrate temperature.

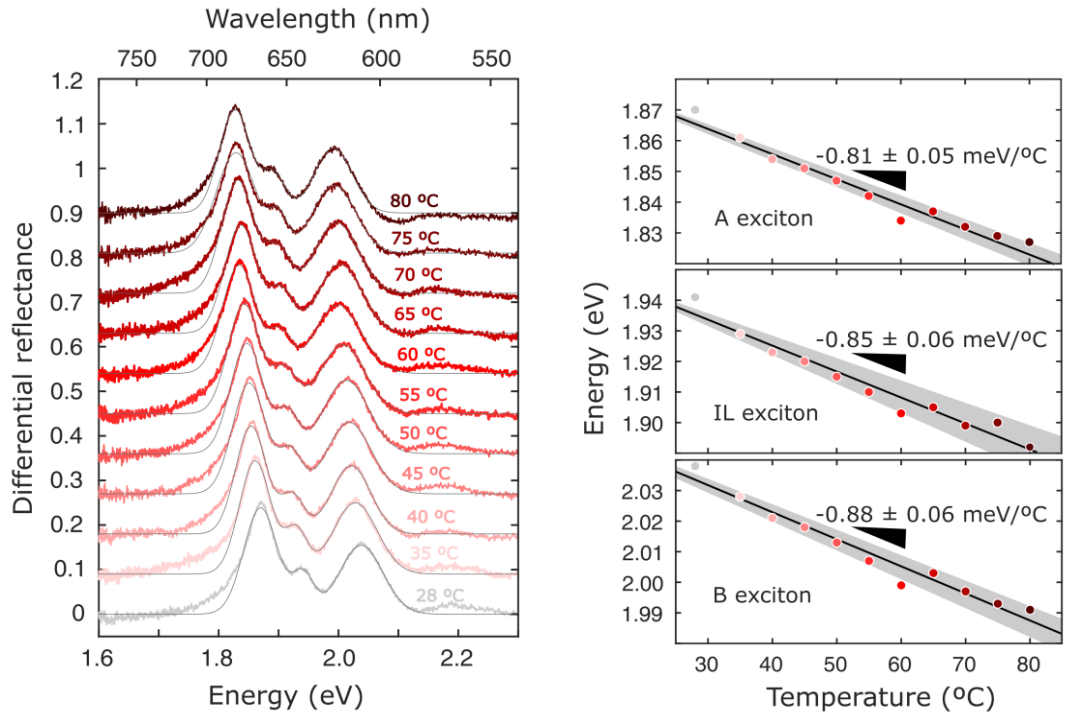

**Figure S8.** Left: differential reflectance spectra of a MoS₂ bilayer deposited on PP recorded at different temperatures. The black solid lines represent the total fit to the data (composed of three Gaussian peaks). Right: energy of the A, B and IL excitonic peaks extracted from the fit plotted as a function of the substrate temperature.

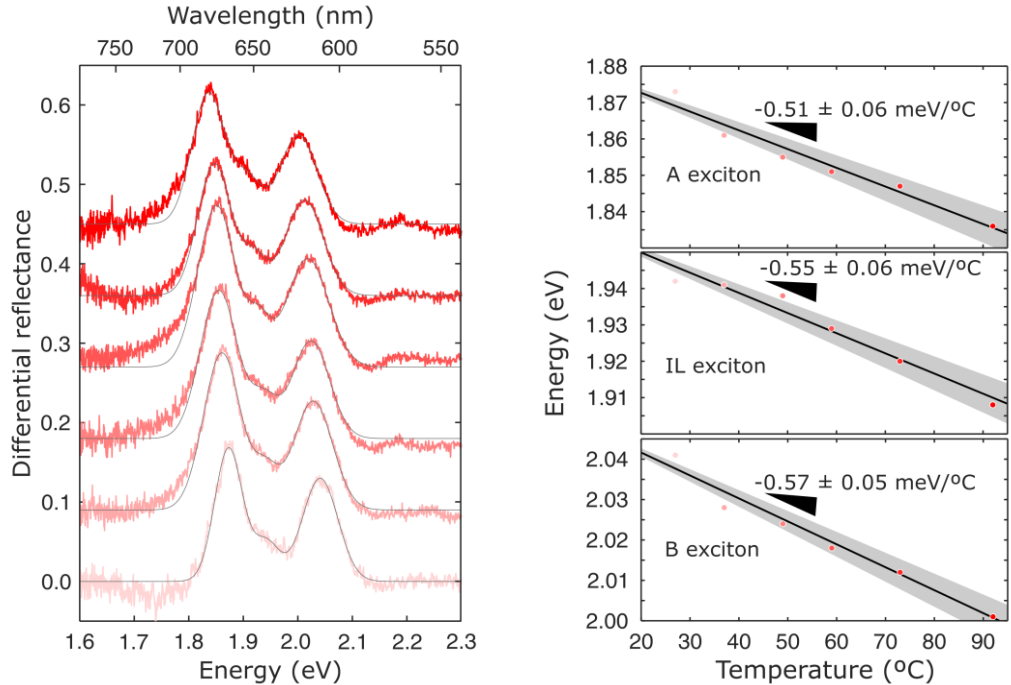

**Figure S9.** Left: differential reflectance spectra of a MoS₂ bilayer deposited on PP recorded at different temperatures. The black solid lines represent the total fit to the data (composed of three Gaussian peaks). Right: energy of the A, B and IL excitonic peaks extracted from the fit plotted as a function of the substrate temperature.

**Trilayer MoS₂**

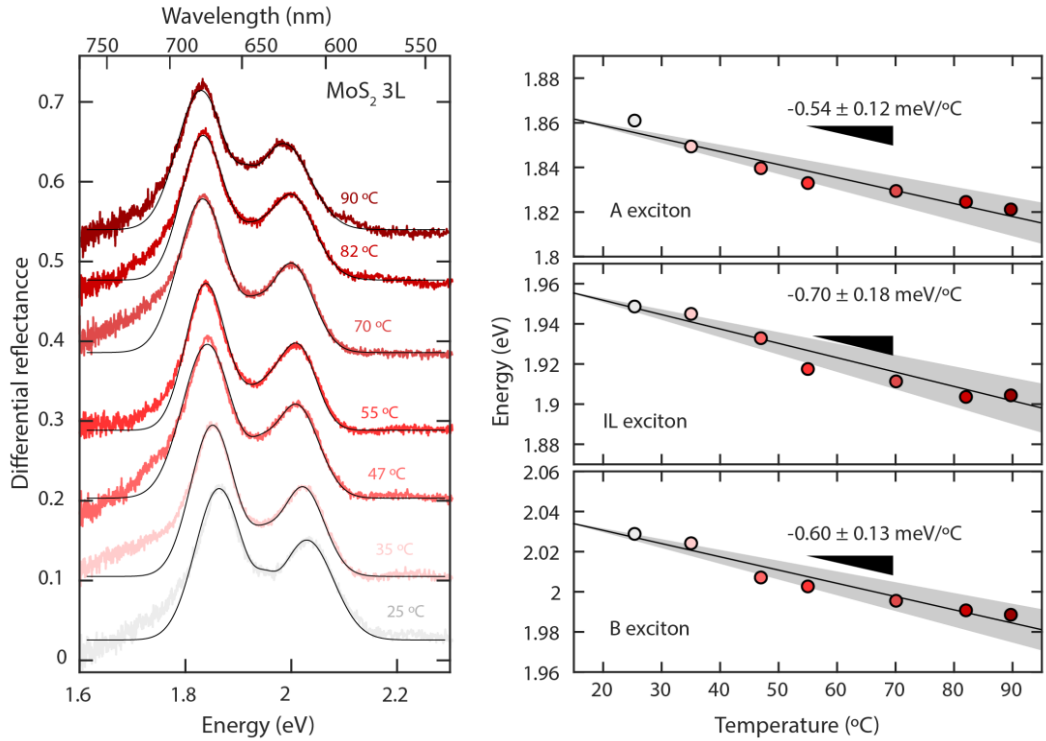

**Figure S10.** Left: differential reflectance spectra of a MoS₂ trilayer deposited on PP recorded at different temperatures. The black solid lines represent the total fit to the data (composed of three Gaussian peaks). Right: energy of the A, B and IL excitonic peaks extracted from the fit plotted as a function of the substrate temperature.

## Exciton gauge factor statistics for bilayer MoS₂

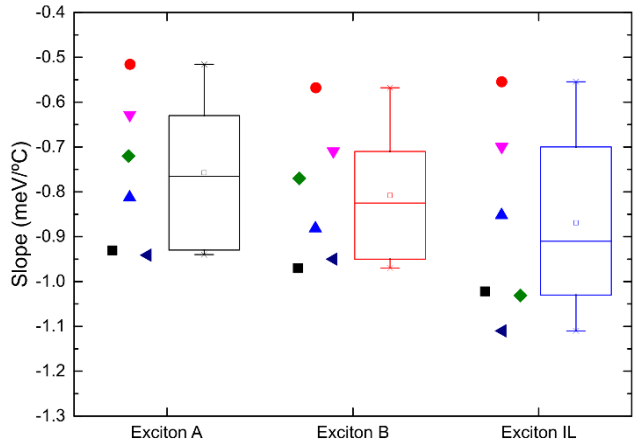

**Figure S11.** Statistics of the experimental gauge factors of the A, B and IL excitons. The filled symbols represent the gauge factors measured for the A, B and IL excitonic peaks in the six bilayer samples studied while the box plots represent the statistical variation. The square points indicate the mean values, the horizontal lines within the boxes the median values, the boxes indicate the quartile values and the vertical lines the maximum and minimum values.

From the quartiles in the box diagrams displayed in Figure S11 we estimate a flake-to-flake statistical fluctuation in the range of 0.3-0.4 meV/°C in the determined gauge factors. Note that even larger flake-to-flake fluctuations have been reported in uniaxial strain experiments.[25] Nonetheless, the difference in gauge factor between the A exciton and the IL and B excitons are substantially different from zero (see Figure S12).

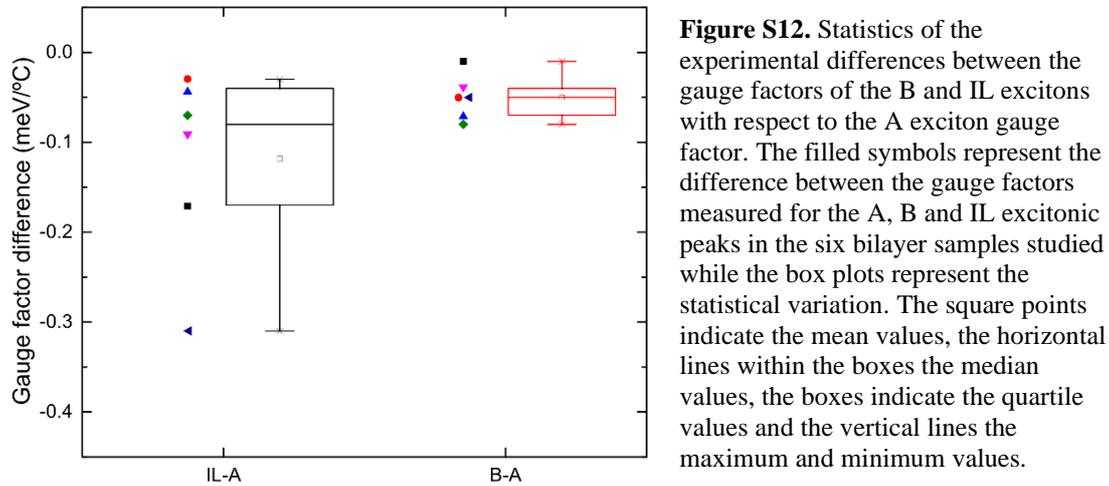

**Figure S12.** Statistics of the experimental differences between the gauge factors of the B and IL excitons with respect to the A exciton gauge factor. The filled symbols represent the difference between the gauge factors measured for the A, B and IL excitonic peaks in the six bilayer samples studied while the box plots represent the statistical variation. The square points indicate the mean values, the horizontal lines within the boxes the median values, the boxes indicate the quartile values and the vertical lines the maximum and minimum values.

Some uniaxial strain experiments attributed the large flake-to-flake to the presence of micro-cracks or other imperfections where the strain is locally released giving rise to a non-uniform strain profile.[16,28] In order to sheed some light in the observed flake-to-flake variation in our experiments we carried out scanning micro-reflectance measurements to probe the spatial variation of the strain gauge in our biaxially strained $MoS_2$ finding a much smaller spatial variation than in uniaxially strained samples which we attribute to a more intimate contact between the flake and the polymer substrate (note that we subject the samples to an initial thermal cycling before starting to measure which might yield to a contact improvement). The variation in the gauge factor due to spatial inhomogeneity can reach 0.1-0.2 meV/ºC. Therefore, there is still another 0.2 meV/ºC variability in the gauge factor whose origin is still unclear.

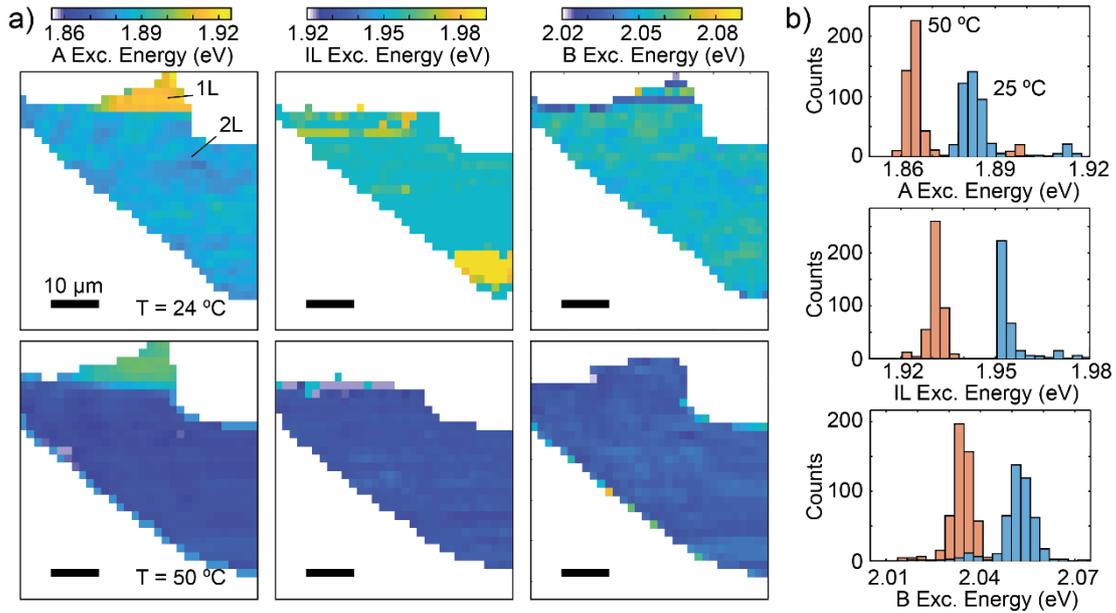

**Figure S13.** (a) Spatial map of the exciton energies at 24°C and 50°C. (b) 1D histograms of the exciton energies at these two temperatures which allows to estimate the spot-to-spot variation in the exciton energy values.

## Section S4 – Disentangling temperature and strain effects

In our experiments the biaxial strain is applied by changing the temperature of the PP substrate therefore we need a way to disentangle the effect arising simply by the temperature change from the effects originated by the biaxial strain. Therefore, we performed a set of measurements on SiO$_2$/Si, a substrate with a negligible thermal expansion ~$1\cdot10^{-6}$ K$^{-1}$ (as compared with the $128\cdot10^{-6}$ K$^{-1}$ of the PP substrate). We selected substrates with 50 nm SiO$_2$ capping layer because they allow to directly resolve the exciton position from differential reflectance measurements. For other SiO$_2$ thicknesses the substrate Fresnel interference could be so strong that hampers the exciton observation.

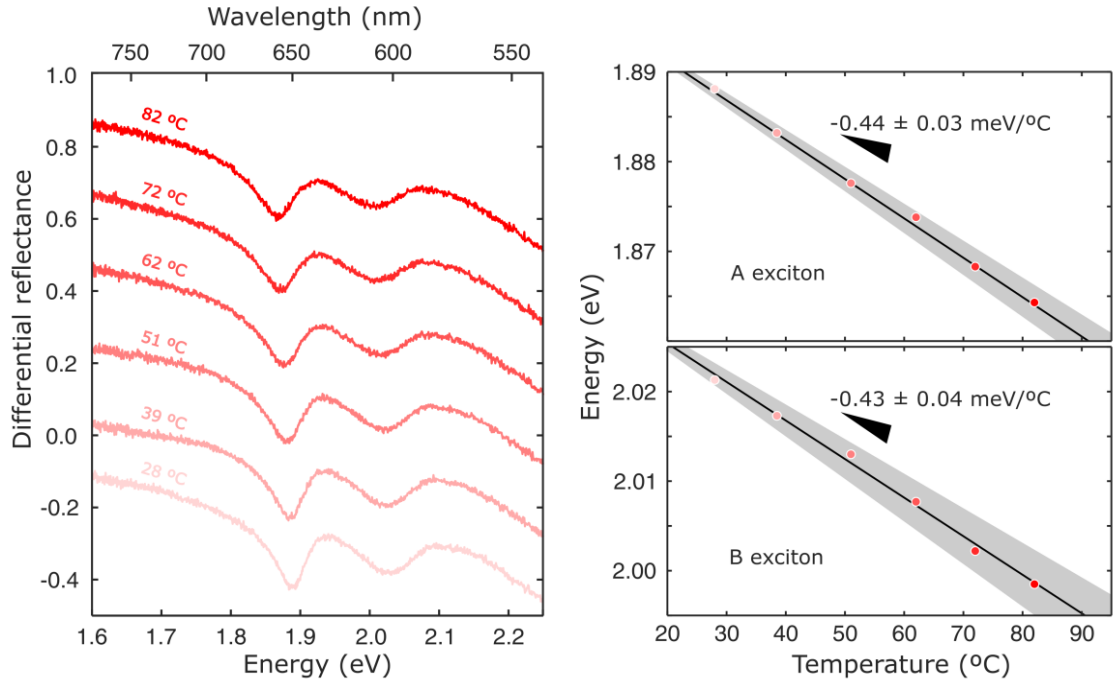

**Figure S14.** Left: differential reflectance spectra of a $MoS_2$ monolayer deposited on 50 nm $SiO_2$/Si recorded at different temperatures. Right: energy of the A and B excitonic peaks plotted as a function of the substrate temperature.

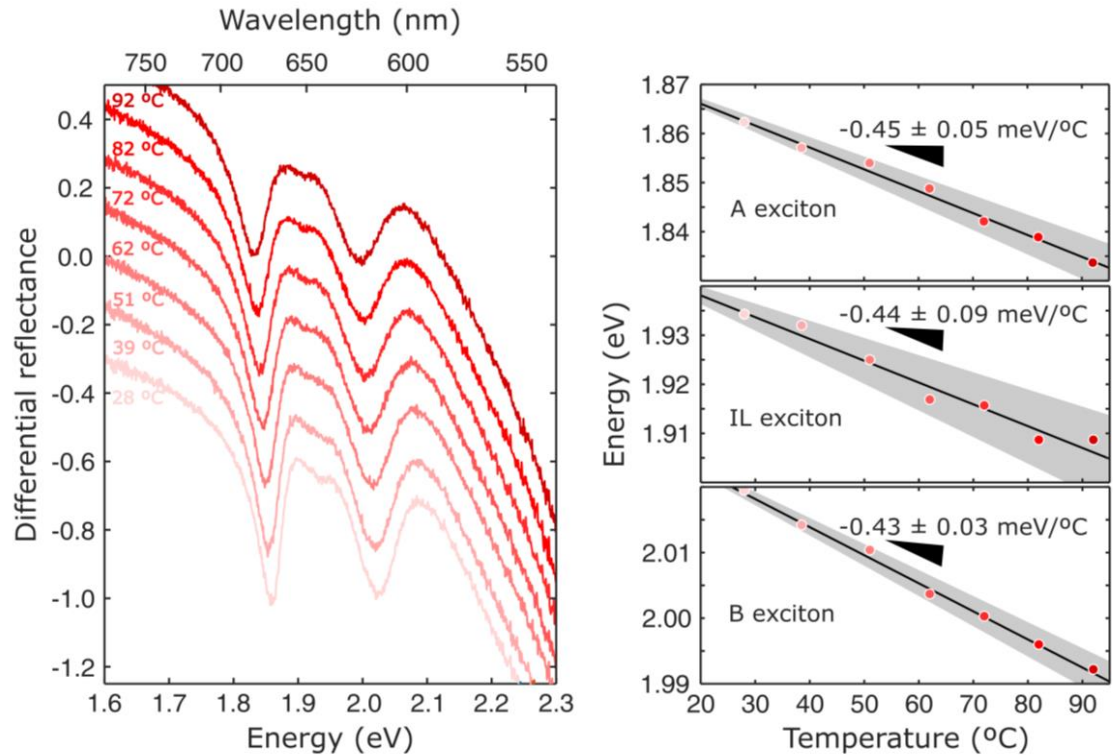

**Figure S15.** Left: differential reflectance spectra of a $MoS_2$ bilayer deposited on 50 nm $SiO_2$/Si recorded at different temperatures. Right: energy of the A, B and IL excitonic peaks plotted as a function of the substrate temperature.

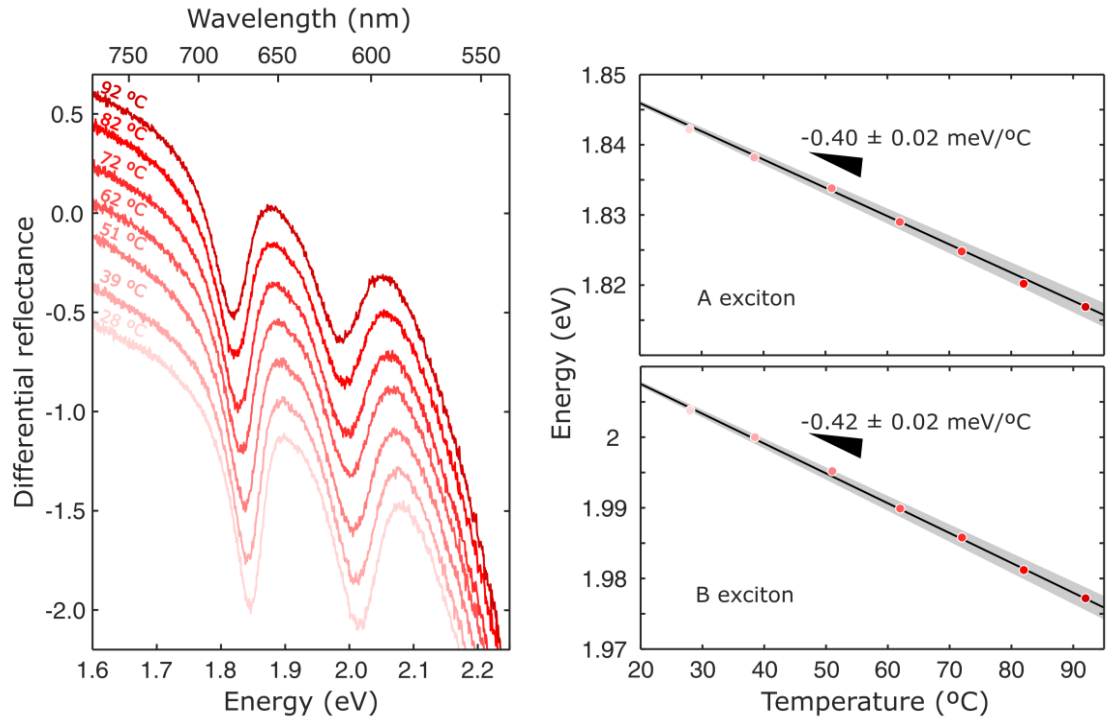

**Figure S16.** Left: differential reflectance spectra of a MoS$_2$ trilayer deposited on 50 nm SiO$_2$/Si recorded at different temperatures. Right: energy of the A and B excitonic peaks plotted as a function of the substrate temperature.

## Section S5 – Ruling out slippage effects

When the flakes are subjected to very large strain they could eventually suffer from slippage or even breakdown. Both events would lead to strain release. We can rule out these two scenarios by checking the reversibility of the strain-release cycles. Figure S15 shows a set of differential reflection spectra acquired in bilayer MoS$_2$ upon a full straining/releasing strain cycle. The vertical gray dotted line indicates the initial position of the A exciton demonstrating the reversibility of the cycle.

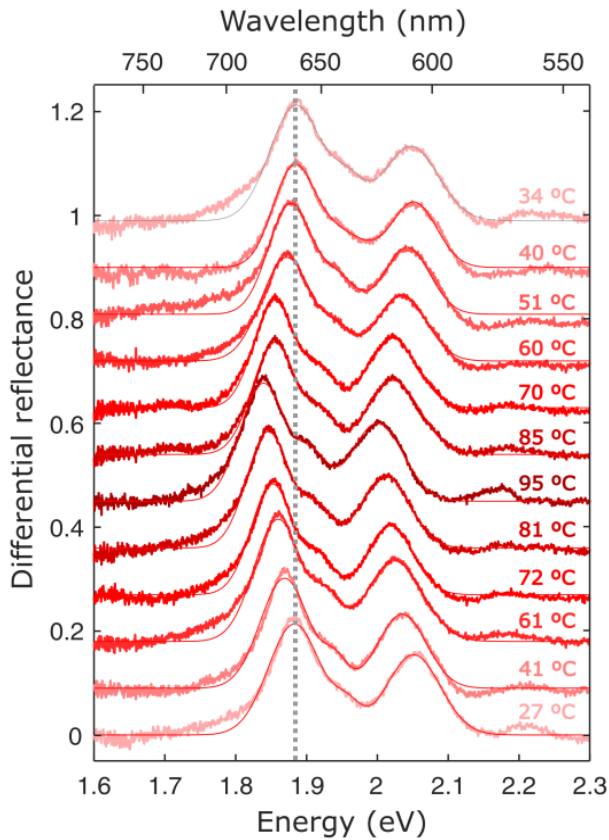

**Figure S17.** Differential reflectance spectra of a MoS$_2$ bilayer deposited on PP recorded at different temperatures during a full straining/releasing strain cycle. The vertical dotted gray line indicates the initial position of the A exciton as a guide to the eye to check the reversibility of the cycle.

## Section S6 – Estimating the strain transfer in multilayered flakes

When the flakes are subjected to very large strain they could eventually suffer from slippage or even breakdown. Both events would lead to strain release. We can rule out these two scenarios by checking the reversibility of the strain-release cycles. Figure S15 shows a set of differential reflection spectra acquired in bilayer MoS$_2$ upon a full straining/releasing strain cycle. The vertical gray dotted line indicates the initial position of the A exciton demonstrating the reversibility of the

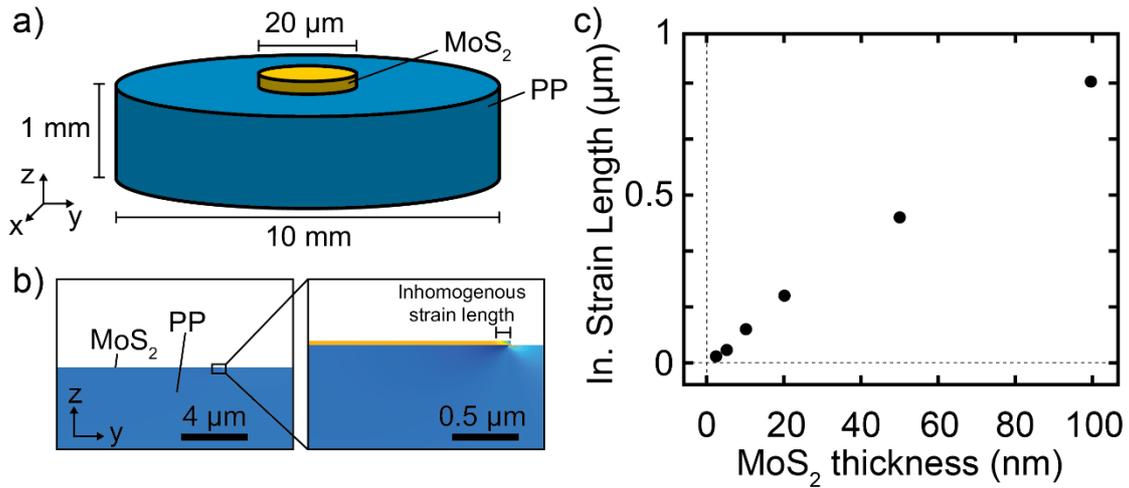

**Figure S18.** (a) Scheme of the 3D mechanical model used for the finite element simulation of the strain transfer. (b) Cross section of the resulting strain transfer. The strain is homogeneous along the thickness of the whole flake. We have simulated flakes up to 100 nm thick. Close to the edge of the flakes there is a region where the strain is not homogeneous, and we found that the length of this region depends on the thickness. This means that if one measure too close to the edge of a very thick multilayered flake the strain would be different than that in the middle of the flake. (c) Relationship between the length of the inhomogeneous strain region and the MoS$_2$ flake thickness.